\documentclass[showpacs,twocolumn,pre,floatfix]{revtex4}
\usepackage{psfrag,epsfig,amsfonts,amssymb,amsmath}
\usepackage{dcolumn}
\usepackage[normalem]{ulem}
\usepackage{color}

\begin{document}
\newcommand{\dout}[1]{\textcolor{red}{\sout{ #1 }}}
\newcommand{\dadd}[1]{\textcolor{green}{ #1 }}

\title{Directing Brownian motion on a periodic surface}

\author{David Speer$^{1}$}
\author{Ralf Eichhorn$^{1,2}$}
\author{Peter Reimann$^{1}$}
\affiliation{$^{1}$Universit\"at Bielefeld, Fakult\"at f\"ur Physik, 33615 Bielefeld, Germany
\\
$^{2}$NORDITA, Roslagstullsbacken 23, 10691 Stockholm, Sweden}

\begin{abstract}
We consider an overdamped Brownian particle, 
exposed to a two-dimensional, square lattice 
potential and a rectangular ac-drive.
Depending on the driving amplitude,
the linear response to a weak dc-force 
along a lattice symmetry axis consist in 
a mobility in basically any direction.
In particular, motion exactly opposite 
to the applied dc-force may arise.
Upon changing the angle of the dc-force
relatively to the square lattice,
the particle motion remains
predominantly opposite to the dc-force.
The basic physical mechanism consists
in a spontaneous symmetry breaking of 
the unbiased deterministic particle dynamics.
\end{abstract}
                                                 
\pacs{05.45.-a, 05.60.-k, 05.40.-a} 

\maketitle

Brownian particle dynamics in two-dimensional 
periodic potential landscapes arise in a large
variety of different contexts.
Examples include driven vortex lattices
\cite{rei99,rei02,ooi07},
surface diffusion \cite{mir05},
a ring of several Josephson junctions \cite{gei96},
colloidal particles or globular DNA in 
structured microfluidic devices \cite{hua04,ros05}
and in optical \cite{kor02,man03} 
or magnetic \cite{tie07} lattices,
enzymatic
reaction cycles
driving molecular motors \cite{mag94},
nanoscale friction \cite{pri03}
and superlubricity \cite{ver04}.
They have recently attracted 
considerable theoretical \cite{rei99,lac05}
and experimental \cite{kor02}
interest for particle sorting
in two-dimensional periodic structures
with the help of an externally applied 
dc-force, whose angle relatively to 
the periodic potential can be 
parametrically changed.
The key point is that the resulting particle 
velocity may exhibit a different direction
than the applied dc-force and that the 
deflection angle may be different for 
different particle species.
While the deflection angles between force
and velocity remain bounded to relatively 
small values, our present system will
lead to (practically) arbitrary 
deflection angles.

A second recent series of papers \cite{rei02,tie07}
considers the same system but in the presence of
an additional circular ac-drive.
Deflection angles up to $90^{\circ}$
(absolute transverse mobility) have been found in
\cite{rei02}, while \cite{tie07} reports transporting orbits
in the absence of a dc-drive.
A related variant is to replace the circular by
a more common, linear ac-drive, 
but now breaking the time-space symmetry by 
chosing a bi-harmonic driving signal \cite{ooi07}, 
and focusing on the low friction regime \cite{gua03}.
The setup we will consider here is related
but simpler: a standard ac-drive without
any concomitant space-time symmetry breaking
and negligible inertia effects.

Unbiased far from equilibrium dynamics 
of a Brownian particle, 
responding to a dc-force by a directed 
transport opposite to that force, have been 
extensively investigated under the label 
``absolute negative mobility'' (ANM)
\cite{ros05,anm}.
Our present work represents the natural extension,
namely an unbiased far from equilibrium system
admitting an easily controllable linear response 
into (practically) any direction relative to the 
dc-force, including ANM as a special case.

\begin{figure}
\epsfxsize=0.6\columnwidth
\epsfbox{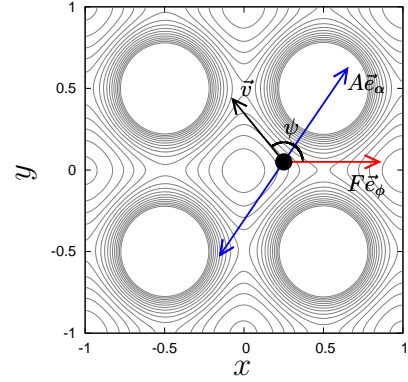} 
\caption{Schematic illustration of our model (\ref{1}).
The contour lines represent the potential $U(\vec{r})$
with a cut-off for better visualization (white discs).
The double arrow indicates the ac-drive $A(t)\vec e_\alpha$,
one arrow the dc-bias $F\vec e_\phi$,
and another arrow the particle velocity (\ref{6}).
The particle is sketched by the black dot.
The angle $\psi$ quantifies the ``deflection'' of
$\vec v$ from $F\vec e_\phi$.
}
\label{fig1}
\end{figure}

We consider the following 2d Langevin dynamics 
of a Brownian particle with coordinates 
$\vec r=x\vec e_x+y\vec e_y$:
\begin{equation}
\dot{\vec r}(t) = A(t)\vec e_\alpha + F\vec e_\phi 
-\nabla U(\vec r(t)) + \sqrt{2\Gamma}\,\vec\xi (t) \ .
\label{1}
\end{equation}
Thus, inertia effects are neglected
(overdamped dynamics),
and the friction coefficient
is absorbed into the time unit.
As illustrated with Fig. \ref{fig1},
$A(t)$ is the ac-driving signal
along the direction
$\vec e_\alpha:=(\cos\alpha,\,\sin\alpha)$,
and analogously for the
dc-bias $F\vec e_\phi$.
The periodic potential is represented
by $U(\vec r)$ and thermal fluctuations
of temperature $\Gamma$ are modeled 
by the two delta-correlated, 
Gaussian noise components of $\vec \xi(t)$.
We focus on the particularly simple rectangular
driving $A(t)= a \, \mbox{sign}\{\cos(\Omega\, t)\}$
with amplitude $a$ and period $T=2\pi/\Omega$.
We verified that a sinusoidal $A(t)$ leaves all our 
main findings qualitatively unchanged and expect 
the same for even more general $A(t)$.
Further, we focus on the potential
$U(\vec r) = \sum \tilde U(\vec r+Ln\vec e_x+Lm\vec e_y)$
with  $r:=|\vec r|$, 
$\tilde U(\vec r) = u\, \exp\{-r/\lambda\}/r$,
and $u,\,\lambda >0$.
In other words, we consider a square lattice 
of repulsive Yukawa potentials, a standard 
choice for screened charges \cite{rei02}.
Again, we expect that our results remain 
qualitatively unchanged for more general
$\tilde U(\vec r)$, modelling e.g. 
pinned vortices \cite{rei99},
optical tweezers \cite{kor02,man03},
or magnetic bubbles \cite{tie07},
and we have explicitly verified this
for Gaussian shaped repulsive and attractive
potentials.
Without loss of generality we choose length
and time units with $L=1$ and $u=1$.
Regarding $\lambda$, we obtained practically
indistinguishable results for all $\lambda\geq 4$,
variations by a few percent down to $\lambda\approx 1$,
and notable quantitative but no qualitative differences
at least down to $\lambda\approx 0.1$.
In the following we focus on the representative 
example $\lambda=4$.

The observable of main interest will be the time-averaged particle
velocity
\begin{equation}
\vec v = v_x\vec e_x+v_y\vec e_y :=
\lim_{t\to\infty} \frac{1}{t} \int_0^\infty dt'\, \dot{\vec r}(t') \ .
\label{6}
\end{equation}
being independent of the seed $\vec r(0)$ and the realization of 
$\vec\xi(t)$ in (\ref{1}) for any $\Gamma>0$ due to ergodicity reasons.

\begin{figure}
\epsfxsize=0.95\columnwidth
\epsfbox{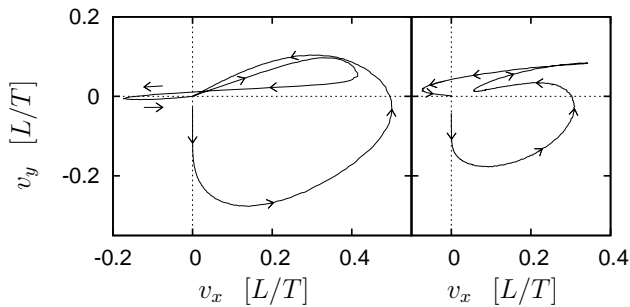} 
\caption{Velocity (\ref{6}) 
from numerical simulations of
(\ref{1}) for $\alpha=0.15\cdot 360^{\circ}=54^{\circ}$, 
$a\in[4.3,\, 8]$, 
$\phi=0^{\circ}$, $F=0.03$, $\Omega=4.3$, 
$\Gamma=2.2 \cdot 10^{-4}$ (left) and 
$\Gamma=6.7 \cdot 10^{-4}$ (right).
Shown is the parametric dependence of the
velocity components $v_x$, $v_y$  
on the ac-amplitude $a$
in units of $L/T$ ($L=1$, $T=2\pi/\Omega$).
Arrows indicate increasing $a$-values.
Upon further decreasing $F$, a close to 
linear response behavior of $\vec v$ 
results (not shown).
}
\label{fig2}
\end{figure}

Generally speaking, the periodic potential, the ac-drive
and the dc-bias in (\ref{1}) give rise to several 
``competing directions'',
whose net effect on the velocity $\vec v$ from
(\ref{6}) is far from obvious.
For zero bias $F$, the ac-forcing still keeps the
system off equilibrium but any non-zero velocity
$\vec v$ is prohibited by symmetries \cite{den08}.
Our first objective is the linear response 
of $\vec v$ to a weak dc-bias along the 
$x$-axis, cf. Fig. \ref{fig1}.
Our findings in Fig. \ref{fig2} exhibit a
quite intriguing behavior. 
Keeping all ``competing directions'' fixed and
solely changing the ac-amplitude by 40\%, 
almost any direction of $\vec v$ may arise,
even motion exactly opposite to the 
applied dc-bias (ANM).

\begin{figure}
\epsfxsize=0.95\columnwidth
\epsfbox{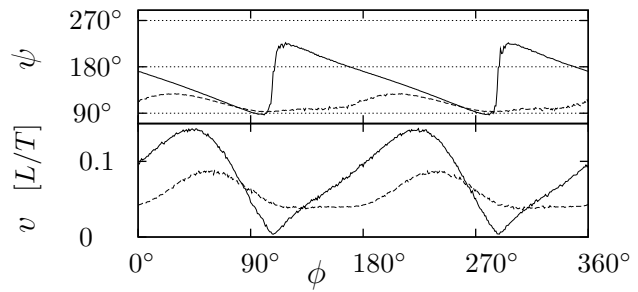} 
\caption{Deflection angle $\psi$ (see Fig. \ref{fig1})
and velocity $v:=|\vec v|$ 
versus ``dc-directionality'' $\phi$
(see (\ref{1})) for the same system as in 
Fig. \ref{fig2}
but with $a=7.4$, 
and $\Gamma=4.4\cdot 10^{-4}$ (solid),
$\Gamma=6.7\cdot 10^{-4}$ (dashed).}
\label{fig3}
\end{figure}

Next, we focus on a set of parameters close to the occurrence 
of ANM in Fig. \ref{fig2} 
and now ask for the response of $\vec v$
upon changing the direction of the dc-bias.
Again, the results in Fig. \ref{fig3} are quite
non-trivial, the most remarkable feature being that
the projection of the velocity along the dc-bias is
mostly negative, i.e. the particle motion remains 
predominantly opposite to the dc-force
($90^\circ< \psi < 270^\circ$).
Similarly as in Fig. \ref{fig2}, the
effect is particularly striking for small noise strengths 
$\Gamma$, and vanishes for $\Gamma\gtrsim 10^{-3}$. 

To better understand these findings
we first focus on the deterministic dynamics 
($\Gamma=0$) in the 
simplest case when the ac-drive and the 
dc-bias are parallel and acting along
one of the main symmetry axes of the 
periodic potential.
Regarding the $(1,0)$ 
direction,
i.e. $\phi = \alpha = 0^\circ$,
the lines $y=n/2$
constitute invariant sets of the deterministic 
dynamics (\ref{1}),
stable for odd and unstable for even $n$.
Thus, the particle motion is confined
between two neighboring such lines and
generically is attracted by the one with 
odd $n$ for large times.
Further, the velocity (\ref{6}) necessarily must 
follow  the direction of the 
dc-force, i.e. $\psi=0^\circ$, 
and, in particular, vanishes for $F=0$.
This qualitative behavior remains unchanged in the 
presence of noise ($\Gamma > 0$).
Analogous conclusions hold for the $(0,1)$ lattice direction.

\begin{figure}
\epsfxsize=0.9\columnwidth
\epsfbox{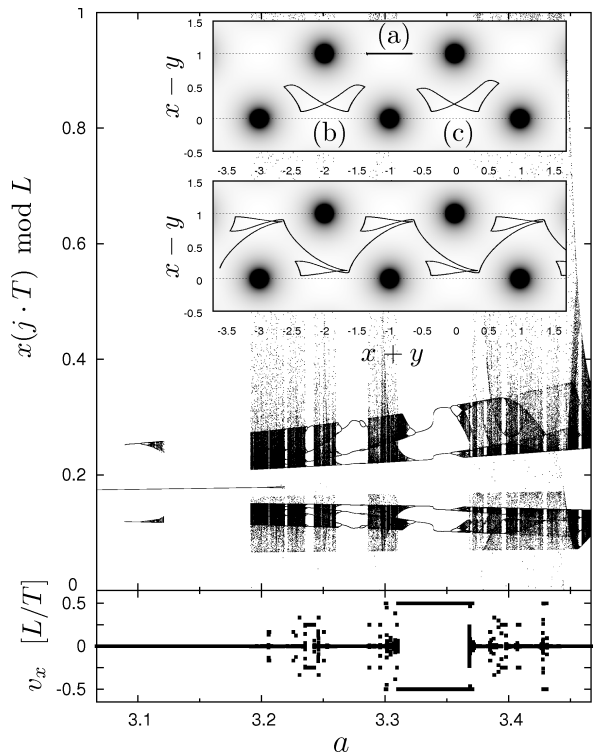} 
\caption{Upper part: Bifurcation diagram 
for the unbiased ($F=0$) deterministic ($\Gamma=0$)
dynamics (\ref{1}) with $\alpha = 45^\circ$,
$\Omega = 3$, and varying $a$.
Shown is a stroboscopic representation of
the attractors, governing the long-time
behavior, by plotting $x(jT)$ modulo $L$ ($L=1$), 
i.e. the reduced $x$-component at multiples $j$ 
of the driving period $T=2\pi/\Omega$,
for several different seeds $\vec r(0)$
after initial transients have died out.
Lower part: Corresponding $x$-component of the
average velocity (\ref{6}) in units of $L/T$.
Upper inset: Stable periodic orbits (attractors)
for $a=3.1$ (a and b) and for $a=3.107$ (c).
Dotted: invariant sets $x-y=n$.
Black ``clouds'' and ``discs'' represent the potential 
$U(\vec{r})$, corresponding to Fig. \ref{fig1}
after a $45^\circ$ rotation.
Lower inset: stable period-2 transporting orbit 
for $a=3.33$.
}
\label{fig4}
\end{figure}

Turning to the $(1,1)$ direction, 
i.e. $\phi = \alpha = 45^\circ$,
the lines $x-y=n$ now constitute invariant sets of the deterministic 
dynamics (\ref{1}) due to its invariance
under $S_1: (x,y) \mapsto (y,x)$.
Again, the particle motion must remain confined
between two adjacent such lines,
implying for the velocity (\ref{6}) that 
$\psi=0^\circ$ and hence $v_x = v_y$.
But now, the motion on the invariant lines
may change its stability properties upon variation 
of a system parameter, and additional
non-trivial attractors, not contained in any of 
the invariant lines, may arise.
A typical example is shown in Fig. \ref{fig4}.
We see that -- depending on the driving amplitude 
$a$ and possibly also on the seed $\vec r(0)$ -- 
the orbit $\vec r(t)$ approaches 
a periodic or a chaotic long time behavior.
The concomitant velocity (\ref{6}) is still well defined 
but -- in contrast to the noisy case $\Gamma>0$ -- 
now may depend on the initial condition $\vec r(0)$.
The ``central'' straight line in the bifurcation diagram for
$a <  3.22$ belongs to all the periodic 
attractors on the invariant sets $x-y=n$, 
exemplified by (a) in the upper inset of Fig. \ref{fig4},
and giving rise to a vanishing average velocity.
At $a \approx 3.09$ we observe the appearance of an additional
pair of non-transporting ($\vec v=0$) periodic attractors, 
spontaneously breaking $S_1$ as well as the second symmetry 
$S_2: (x,y) \mapsto (-x,-y)$ of (\ref{1}), but still maintaining 
(up to translations) $S_1 \circ S_2: (x,y) \mapsto (-y,-x)$,
see (b) in Fig. \ref{fig4}.
Thus, there are now three coexisting attractors
within every unit cell of the periodic potential,
one of type (a), the others of type (b) 
and its image under $S_1$.
At $a \approx 3.102$ the pair of type (b) attractors
exhibits a pitchfork bifurcation, which spontaneously breaks
the $S_1\circ S_2$ symmetry (symmetry breaking bifurcation \cite{swi84}),
resulting in four distinct, 
non-transporting attractors 
per unit cell. One of them is exemplified with (c) in Fig. \ref{fig4},
its three ``brothers'' follow as mirror images
with respect to the closest $x-y=n$ and/or 
$x+y=n$ lines. Upon further increasing $a$, 
a period doubling route to chaos follows, 
which would be impossible without the preceding
symmetry breaking bifurcation \cite{swi84}.
The corresponding attractors loose stability by way 
of a crisis at $a \approx 3.12$. 
They reappear beyond $a \approx 3.19$ as ``chaotic bands'',
interrupted by ``periodic windows''.
Some of these windows exhibit attractors 
corresponding to phase-locked \emph{transporting} 
orbits. 
The symmetry breaking bifurcation at $a\approx 3.102$ 
is pivotal for such transporting orbits:
Since there is no systematic force ($F=0$), 
which could favor motion in one or the other direction,
{\em spontaneous symmetry breaking} of all 
symmetries involving $S_2$ is an indispensable
prerequisite for transporting orbits.
The simplest and most prominent example arises 
within the periodic window at 
$a\approx 3.33$, exhibiting two attractors
between any pair of adjacent $x-y=n$ lines.
One such orbit is exemplified
in the lower inset of Fig. \ref{fig4}, its ``twin brother''
follows as mirror image with respect to any $x+y=n$ line.
Apparently, this orbit arises 
by continuing the deformation of orbit (b), which leads 
to (c) even further, and rewiring one of its
``arms'' into the neighboring unit cell.
The latter operation cannot be realized
by a continuous deformation and hence one might guess
that this somehow happens within the ``gap''
in the bifurcation diagram between 
$a \approx 3.12$ and $a \approx 3.19$. 

Due to the $S_1\circ S_2$ symmetry at $F=0$,
oppositely transporting orbits co-exist and are stable
within exactly the same range of the other system parameters.
Applying a force $F \neq 0$, however, breaks the 
symmetry and hence the existence regions of the two orbits 
in parameter space no longer coincide.
Closer inspection of how these regions change 
shape and size reveals that there are, for not too large $|F|$, 
parameter values, where \emph{the only stable orbit 
is the one which transports against the dc-bias $F$}.
In other words, we recover yet another example
of ``pure'' ANM ($\psi=180^\circ$) \cite{ros05,anm},
which furthermore turns out to survive even 
in the presence of (sufficiently weak) noise 
($\Gamma>0$).

We finally address the case of arbitrary 
(but fixed) orientations of the external forcings.
Without loss of generality we restrict ourselves to
$0 < \alpha < 90^\circ$ but admit arbitrary $\phi$.
For such general driving directions
the deterministic dynamics (\ref{1}) is not 
restricted any more by simple invariant sets. 
Rather, by means of extensive computer simulations we have 
found that transporting particle motion
is created in a way very similar to the case exemplified 
above with Fig. \ref{fig4}, and typically 
``locks'' to one of the three main symmetry
axes $(1,0)$, $(0,1)$ or $(1,1)$, 
depending on amplitude and frequency of
the ac-drive.
In particular, for $F=0$, symmetry dictates the
\emph{co-existence of transport into opposite directions}
for either of these basic orientations.
Applying a (not too large)
dc-bias $F\vec e_\phi$ along a direction that is generally 
different from the one of the ac-drive, $\phi \neq \alpha$, has two main effects:
First, this co-existence is lifted, yielding parameter
regions where only one transporting direction out of the 
$6$ different possible directions is stable.
Second, new transport directions around the orientation of the
bias force become accessible that
follow a similar ``locking scheme'' as in \cite{lac05}
upon variation of a system parameter.

In other words, systematically changing, say, the
driving amplitude $a$ leads to ``jumps'' in the deterministic transport direction 
where -- due to the time-dependent ac-drive -- also transport
with velocity components opposite to the bias force occurs, 
in marked contrast to previous findings in \cite{lac05}.
The main effect of (weak) thermal noise is to ``interpolate'' between
neighboring deterministic directions, resulting in the smooth 
behavior shown in Fig. \ref{fig2}: transport in virtually any
direction 
but with emphasis on directions around the orientation 
of the dc-bias.
Similarly, the variations of the deflection angle $\psi$ 
observed in Fig. \ref{fig3}
result from a basically unchanged orientation of 
$\vec v$ into the ``negative'' $(1,0)$
direction as long as the dc-bias has a
non-vanishing component in the ``positive'' $(1,0)$ direction
(i.e. $-90^\circ \lesssim \phi \lesssim 90^\circ$)
and a quick transition into the opposite situation
when the dc-orientation $\phi$ moves into the
complementary regime between $90^\circ$ and  $270^\circ$.
In any case, too large noise strengths $\Gamma$ basically override
the effects of the periodic potential and the system tends to
return to the trivial behavior in the absence of the
periodic potential.

In conclusion, several quite astonishing linear
response transport phenomena of a very simple
and general non-equilibrium system have been observed
and understood to the extent that an efficient
and systematic search of pertinent parameter 
regions becomes easily feasible.
The system is minimal in so far as any further
reduction or simplification unavoidably
rules out one of the indispensable prerequisites,
most notably the occurrence of spontaneous symmetry
breaking and chaos in the deterministic limit.
On the other hand, the effects are robust against
a large variety of modifications and amendments 
of the model and hence should be realizable
in several different experimental systems 
\cite{gei96,hua04,ros05,kor02,man03,tie07}
for instance for particle sorting purposes.

\begin{center}
\vspace{-5mm}
---------------------------
\vspace{-4mm}
\end{center}
We thank J.~Nagel, D.~Koelle, and R.~Kleiner 
for fruitful discussions.
This work was supported by Deutsche Forschungsgemeinschaft 
under SFB 613 and RE1344/4-1

\end{document}